\documentclass{appolb}
\usepackage{graphicx}
\usepackage{amssymb}   

\begin{document}

\title{Remarks on heavy-quark production \\ in photon-nucleon
and photon-photon collisions }

   \author{A. Szczurek$^{1},^{2}$  \\
   {\it $^{1}$ Institute of Nuclear Physics, PL-31-342 Cracow, Poland  } \\
   {\it $^{2}$ University of Rzesz\'ow, PL-35-959 Rzesz\'ow, Poland } \\ }
\maketitle

\begin{abstract}
I discuss mechanisms of heavy quark production
in (real) photon-nucleon and (real) photon - (real) photon
collisions. In particular, I focuse on application of
the Saturation Model.
In addition to the main dipole-nucleon or dipole-dipole contribution
included
in recent analyses, I propose how to calculate within
the same formalism the hadronic single-resolved contribution
to heavy quark production.
At high photon-photon energies this yields a sizeable correction of about
30-40 \% for inclusive charm production and 15-20 \%
for bottom production.
Adding all possible contributions to
$e^+ e^- \rightarrow b \bar b X$ together removes a huge deficit
observed in earlier works but does not solve the problem totally.
\end{abstract}

\section{Introduction}

The total cross section for virtual photon - proton
scattering in the region of small $x$ and intermediate $Q^2$
can be well described by the Saturation Model (SAT-MOD) 
\cite{GW98}.
The very good agreement with experimental data can be extended
even to the region of rather small $Q^2$ by adjusting
an effective quark mass.
At present there is no deep understanding of the fit value
of the parameter as we do not understand in detail the confinement
and the underlying nonperturbative effects related to large size
QCD contributions.

In this presentation I shall
limit to the production of heavy quarks which is
simpler and more transparent for real photons.
Here one can partially avoid the problem of the poor understanding
of the effective light quark mass, i.e. the domain of the large
(transverse) size of the hadronic system emerging from the photon.

It was shown recently that the simple SAT-MOD description
can be succesfully extended also to the photon-photon
scattering \cite{TKM01}.
The heavy quark production in photon-photon collisions is interesting
in the context
of a deficit of standard QCD predictions relative
to the experimental data as observed recently for $b$ quark
production.

\section{Heavy quark production in photon-nucleon scattering}

In the so-called dipole picture the cross section for
heavy quark-antiquark ($Q \bar Q$) photoproduction on
the nucleon can be written as
\begin{equation}
\sigma_{\gamma N \rightarrow Q \bar Q}(W) =
\int d^2 \rho d z \; |\Phi_T^{Q \bar Q}(\vec{\rho},z)|^2
\sigma_{d N}(\rho,z,W) \; ,
\label{SM_gamma_nuc}
\end{equation}
where $\Phi_T$ is (transverse) quark-antiquark photon wave
function (see for instance \cite{NZ91}) and
$\sigma_{dN}$ is the dipole-nucleon total cross section.
Inspired by its phenomenological success
\cite{GW98} we shall use the SAT-MOD parametrization
for $\sigma_{dN}$. 
Because for real photoproduction the Bjorken-$x$ is not defined
we are forced to replace $x$ by $x_g$
\cite{szczurek}.

In Fig.1a I show predictions of SAT-MOD for charm photoproduction.
The dotted line represents calculations based on Eq.(\ref{SM_gamma_nuc}).
The result of this calculation exceeds considerably
the fixed target experimental data.
One should remember, however, that the simple formula (\ref{SM_gamma_nuc})
applies at high energies only. At lower energies one should
include effects due to kinematical threshold.
In the momentum representation this can be done by requiring:
$M_{Q \bar Q} < W$, where $M_{Q \bar Q}$ is the invariant mass of the
final $Q \bar Q$ system.
This upper limit
still exceeds the low energy experimental data.
There are phase space limitations in the region
$x_g \rightarrow$ 1 which have been neglected so far. 
Those can be estimated using naive counting rules.
Such a procedure leads to a reasonable agreement with the
fixed target experimental data.

The deviation of the solid line from the dotted line gives an idea
of the range of the safe applicability of SAT-MOD for
the production of the charm quarks/antiquarks. 
The cross section for $W >$ 20 GeV is practically independent
of the approximate treatment of the threshold effects.
SAT-MOD seems to slightly underestimate the H1 collaboration
data \cite{H1_ccbar}.
For comparison in Fig.1 we show the result of
similar calculations in the collinear approach (thick dash-dotted line)
with details described in \cite{szczurek}.

The calculation above is not complete.
For real photons a vector dominance contribution due to
photon fluctuation into vector mesons should
be included on the top of the dipole contribution.
In the present calculation we include only the dominant
gluon-gluon fusion component. Then
\begin{equation}
\sigma_{\gamma N \rightarrow Q \bar Q}^{VDM}(W) =
\sum_V \frac{4 \pi}{f_V^2} \; \int dx_V dx_N \;
g_V(x_V,\mu^2_F) \; g_N(x_N,\mu^2_F) \; \sigma_{gg \rightarrow Q \bar
  Q}(\hat W) \; .
\label{gp_VDM}
\end{equation}
Here the $f_V$ constants describe the transition of the photon
into vector mesons ($\rho$, $\omega$, $\phi$).
The gluon distributions in vector mesons are taken as that
for the pion \cite{GRV_pion}.

The dash-dotted line in Fig.1a shows the VDM contribution
calculated in the leading order (LO) approximation for
$\sigma_{gg \rightarrow Q \bar Q}$.
The so-calculated VDM contribution is small at small energies
but cannot be neglected at high energies.

The situation for bottom photoproduction seems similar.
In Fig.1b I compare the SAT-MOD predictions
with the data from the H1 collaboration \cite{H1_bbbar}.
Here the threshold effects survive up to very high energy
$W \sim$ 50 GeV.
Again the predictions of SAT-MOD are slightly below
the H1 experimental data point. The relative magnitude
of the VDM component is similar as for the charm production. 

\vspace{-0.1cm}
\begin{figure}
  \begin{center}
\hspace{-1.4cm}
    \includegraphics[width=5.9cm]{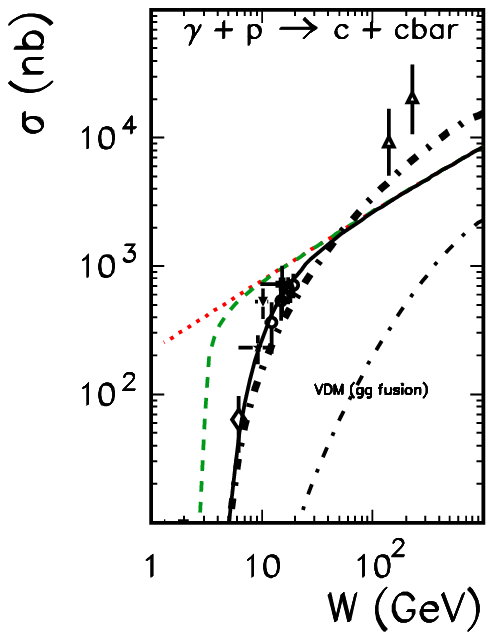}
\hspace{0.0cm}
    \includegraphics[width=5.9cm]{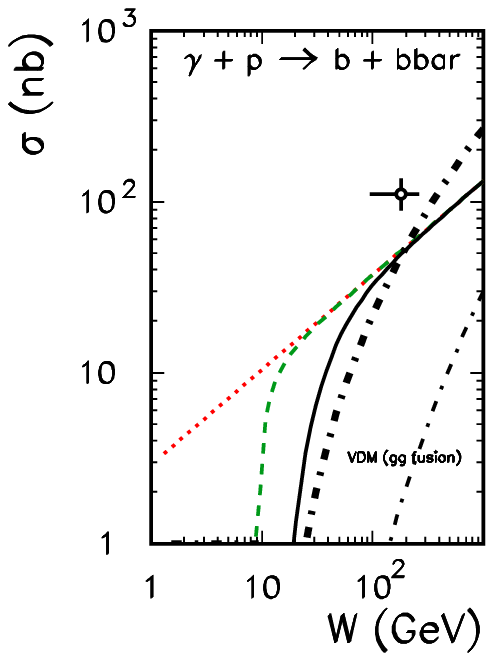}
  \end{center}
\vspace{-1.3cm}
\caption{\small 
The cross section for $\gamma + p \rightarrow Q \bar Q X$.
The dotted line: standard SAT-MOD,
the dashed line: includes kinematical threshold,
the solid line includes in addition a suppression by 
$(1-x_c)^7$, the thick dash-dotted line:
collinear approximation and the thin dash-dotted line:
the LO VDM contribution. 
}
\label{figure1}
\end{figure}

\section{Heavy quark production in photon-photon scattering}

In the dipole-dipole approach 
the total cross section for $\gamma \gamma \rightarrow Q \bar Q$
production can be expressed as
\begin{eqnarray}
\sigma_{\gamma \gamma \rightarrow Q \bar Q}^{dd}(W) &=&
\sum_{f_2 \ne Q} \int
                 |\Phi^{Q \bar Q}(\rho_1, z_1)|^2
                 |\Phi^{f_2 \bar f_2}(\rho_2, z_2)|^2
         \sigma_{dd}(\rho_1,\rho_2,x_{Qf}) \;
d^2 \rho_1 d z_1 d^2 \rho_2 d z_2
\nonumber \\
 &+&
\sum_{f_1 \ne Q} \int
                 |\Phi^{f_1 \bar f_1}(\rho_1, z_1)|^2
                 |\Phi^{Q \bar Q}(\rho_2, z_2)|^2
         \sigma_{dd}(\rho_1,\rho_2,x_{fQ}) \;
d^2 \rho_1 d z_1 d^2 \rho_2 d z_2 \; ,
\nonumber \\
\label{SM_QQbar}
\end{eqnarray}
where
$\sigma_{dd}$ is the dipole-dipole cross section.

There are two problems associated
with direct use of (\ref{SM_QQbar}).
First of all, it is not completely clear how to generalize
$\sigma_{dd}$ from $\sigma_{dN}$
parametrized in \cite{GW98}.
Secondly, formula (\ref{SM_QQbar}) is correct only at
$W \gg 2 m_Q$. At lower energies one should worry about
proximity of the kinematical threshold.

In a very recent paper \cite{TKM01} a new phenomenological
parametrization for 
$\sigma_{dd}$ has been proposed.
The phenomenological threshold factor in \cite{TKM01}
does not guarantee automatic vanishing of the cross section
exactly below the true kinematical threshold $W = 2 m_a + 2 m_b$.
Therefore, instead of the phenomenological
factor I impose an extra kinematical constraint:
$M_{f \bar f} + M_{Q \bar Q} < W$ on the integration
in (\ref{SM_QQbar}).

It is also not completely clear how to generalize the energy
dependence of $\sigma_{d N}$ in photon-nucleon scattering
to the energy dependence of $\sigma_{dd}$ in photon-photon
scattering. 
In \cite{szczurek} I have defined the parameter which
controls the SAT-MOD energy dependence
of $\sigma_{dd}$ in a symmetric way 
with respect to both photons.
In comparison to the prescription in \cite{TKM01}, our prescription leads
to a small reduction of the cross section far from the threshold
\cite{szczurek}.

Up to now we have calculated the contribution when
photons fluctuate into quark-antiquark pairs, which is not complete.
The dipole approach must be supplemented by the resolved photon
contribution. This contribution can be estimated in the vector
dominance model when either of the photons fluctuates into vector mesons.
If the first photon fluctuates
into the vector mesons, the so-defined single-resolved contribution
to the heavy quark-antiquark production can be calculated
analogously to the photon-nucleon case as
\begin{equation}
\sigma^{SR,1}_{\gamma \gamma \rightarrow Q \bar Q}(W) =
\sum_{V_1} \frac{4 \pi}{f_{V_1}^2} \int
|\Phi_2^{Q \bar Q}(\rho_2,z_2)|^2 \sigma_{V_1 d}(\rho_2,x_1) \;
d^2 \rho_2 dz_2
\; ,
\label{SR_1}
\end{equation}
%
where $\sigma_{V_1 d}$ is vector meson - dipole total cross section.
In the spirit of SAT-MOD, we parametrize the latter
exactly as for the photon-nucleon case \cite{GW98}
with a simple rescaling of the normalization factor
$\sigma_0^{dV} = 2/3 \cdot \sigma_0^{dN}$.
In the present calculation $\sigma_0^{dN}$ as well as
the other parameters of SAT-MOD are taken from \cite{GW98}.
Analogously, if the second photon fluctuates into vector mesons
we obtain
\begin{equation}
\sigma^{SR,2}_{\gamma \gamma \rightarrow Q \bar Q}(W) =
\sum_{V_2} \frac{4 \pi}{f_{V_2}^2} \int
|\Phi_1^{Q \bar Q}(\rho_1,z_1)|^2 \sigma_{d V_2}(\rho_1,x_2) \;
d^2 \rho_1 dz_1
\; .
\label{SR_2}
\end{equation}
This clearly doubles the first contribution (\ref{SR_1}) to the total
cross section.

The integrations in (\ref{SR_1}) and (\ref{SR_2}) are not free
of kinematical constraints.
When calculating both single-resolved contributions,
it should be checked additionally if
the heavy quark-antiquark invariant mass $M_{Q \bar Q}$ is
smaller than the total photon-photon energy $W$ (see \cite{szczurek}).

In Fig.2 I show different contributions to the
inclusive $c / \bar c$ (left panel) and $b / \bar b$ (right panel)
production in photon-photon scattering.
The thick solid line represents the sum of all contributions.

Let us start from the discussion of the inclusive charm production.
The experimental data of the L3 collaboration \cite{L3_gg_ccbar}
are shown for comparison. The modifications discussed above
lead to a small damping of the cross section in comparison
to Ref.\cite{TKM01}. The corresponding result (long-dashed line)
stays below the recent experimental data of the L3 collaboration
\cite{L3_gg_ccbar}. The hadronic single-resolved contribution
constitutes about 30 - 40 \% of the main SM contribution.
At high energies the cross section for the $2c 2\bar c$ 
component is about 8 \% of that for the single $c \bar c$ pair
component. In the inclusive
cross section its contribution should be doubled because
each of the heavy quarks/antiquarks can be potentially identified
experimentally.

\vspace{-0.1cm}
\begin{figure}
  \begin{center}
\hspace{-1.4cm}
    \includegraphics[width=5.9cm]{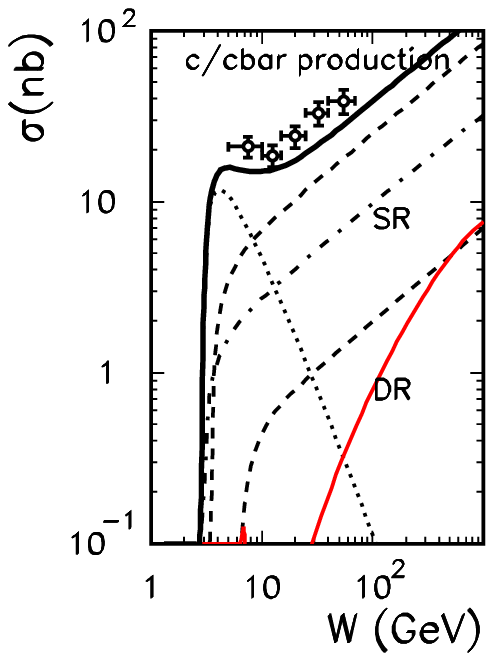}
\hspace{0.0cm}
    \includegraphics[width=5.9cm]{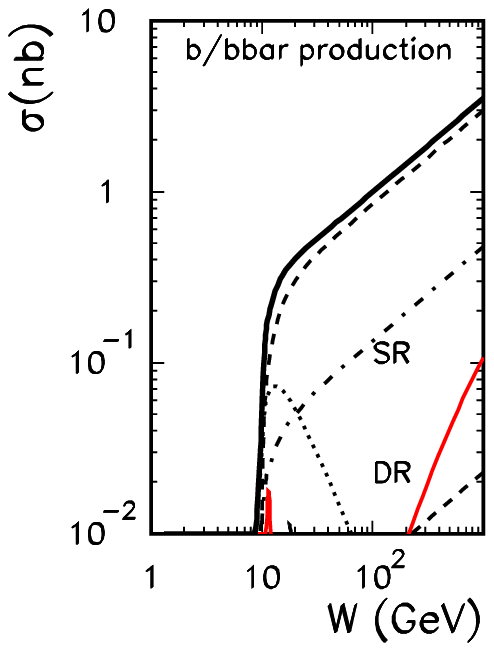}
  \end{center}
\vspace{-1.3cm}
\caption{\small Different contributions to the inclusive charm (left panel)
and bottom (right panel) production.
The long-dashed line:
the dipole-dipole contribution, the dash-dotted line:
the single-resolved contribution,
the lower dashed line: the $2Q  2\bar Q$
contribution, the dotted line: the direct
contribution, the gray solid line: double-resolved contribution.
The experimental data for inclusive $c/\bar c$ production are from
Ref.\cite{L3_gg_ccbar}.
}
\label{figure2}
\end{figure}

At higher energies the direct contribution
is practically negligible.
In contrast, the hadronic double-resolved contribution,
when each of the two
photons fluctuates into a vector meson \cite{szczurek}
is shown by the thin solid line in the figure
becomes important only at very high energies relevant for TESLA.
Here we have consistently taken $g_V(x_V,\mu_F^2) =
g_{\pi}(x_V,\mu_F^2)$.

The situation for bottom production (see right panel) is somewhat
different.
Here the main SAT-MOD component is dominant. Due to smaller charge
of the bottom quark than that for the charm quark, the direct component
is effectively reduced with respect to the dominant SAT-MOD component
by the corresponding ratio of quark/antiquark
charges: $(1/9)^2 : (4/9)^2 = 1/16$.
The same is true for the $2 b 2 \bar b$ component.
Here, in addition,
there are threshold effects which play a role up to relatively
high energy.

\subsection{Short- versus long-distance phenomena}

\begin{figure}
  \vspace{-1cm}
  \begin{center}
    \hspace{-1.2cm}
      \includegraphics[width=6.0cm]{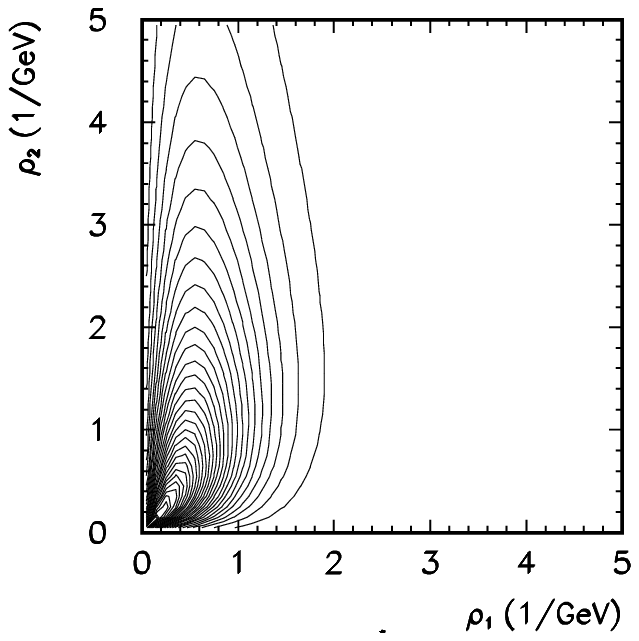}
    \hspace{0.0cm}
      \includegraphics[width=6.0cm]{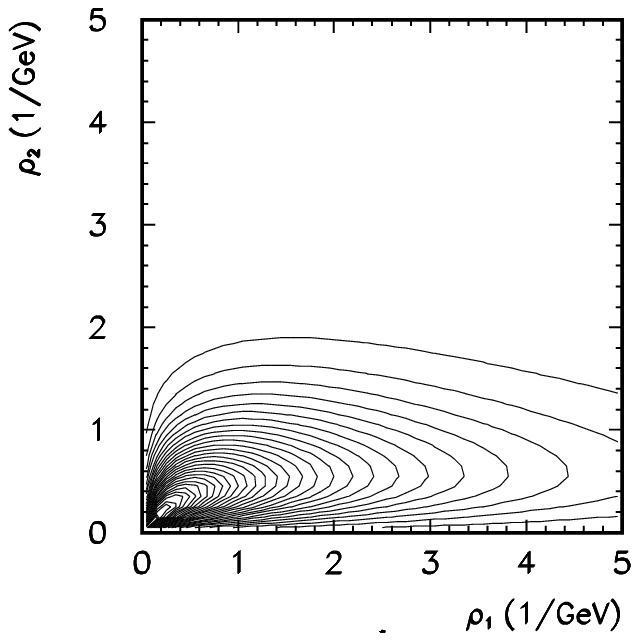} \\ 
    \hspace{-1.2cm}
      \includegraphics[width=6.0cm]{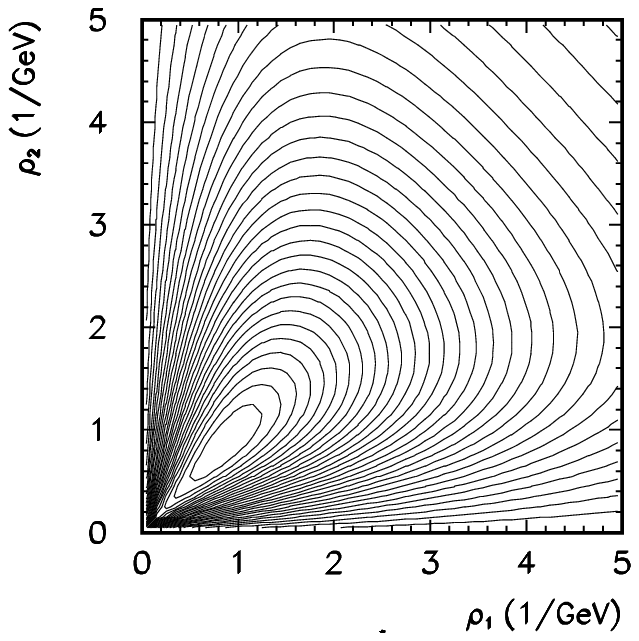}
    \hspace{0.0cm}
      \includegraphics[width=6.0cm]{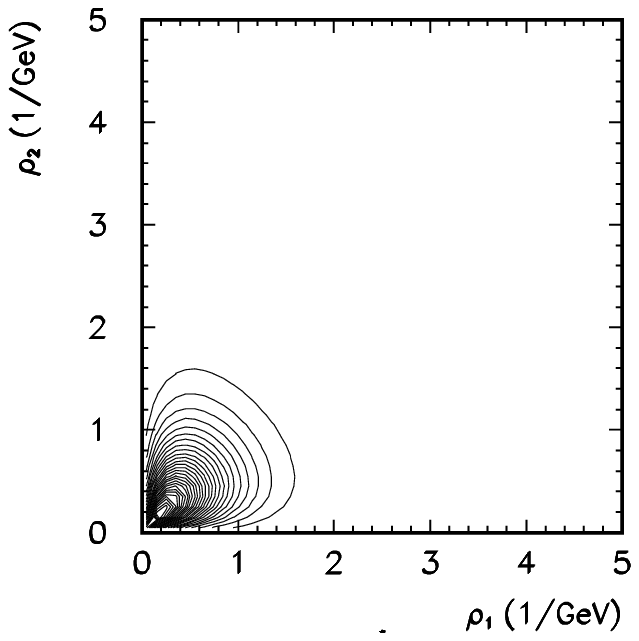}
    \caption{\small A map of
      $\frac{d^2 \sigma^{\gamma \gamma \rightarrow c \bar c}(\rho_1,\rho_2)}
      {d \rho_1 d \rho_2}$
      at W = 100 GeV for the first (left-top panel) and
      the second (right-top panel) photon fluctuating
      into $c \bar c$. For comparison I show analogous maps for light
      quark-antiquark pairs (left-bottom panel) and for
      the case when both pairs consist of charm quarks/antiquarks
      (right-bottom panel).}
    \label{map_rho1_rho2}
  \end{center}
\end{figure}

What are typical distances probed in heavy quark production ?
Is the heavy quark production a short distance phenomenon ?
These questions can be easily answered in the mixed representation
formulation considered in the present paper.
In Fig.\ref{map_rho1_rho2} I display the integrand of
\begin{equation}
\sigma_{\gamma \gamma \rightarrow c \bar c}(W) =
\int \; I(\rho_1,\rho_2) \; d \rho_1 d \rho_2 \; .
\end{equation}
The maxima in Fig.\ref{map_rho1_rho2}
correspond to the most probable situations.
For one light ($m_u$ = $m_d$ = $m_0$, $m_s$ = $m_0$ + 0.15 GeV)
and second heavy quark-antiquark pair the map is clearly asymmetric.
One can observe a ridge parallel to
the $\rho_1$ or $\rho_2$ axis. There is no well localized
maximum. Both short and long distances are probed.

For comparison, in the bottom part of the figure,
I show similar maps when both pairs consist
of light (u,d,s) quarks/antiquarks (left-bottom) and
in the case when both pairs consist of charm quarks/antiquarks
(right-bottom). For light quarks (u,d,s) one observes a
clear maximum at $(\rho_1,\rho_2)$ = (1 GeV$^{-1}$, 1 GeV$^{-1}$)
= (0.2 fm, 0.2 fm). In this case a non-negligible
strength extends, however, up to large distances $\rho_1$ and
$\rho_2$.
Only in the case of the production of two $c \bar c$ pairs,
the cross section is dominated exclusively by short-distance
phenomena.

\subsection{$e^+ e^- \rightarrow e^+ e^- b \bar b X$}

Up to now no attempt was done to unfold experimentally
the cross section for
$\gamma \gamma \rightarrow b \bar b X$.
Only the cross section for the $e^+ e^- \rightarrow b \bar b X$ reaction
was obtained recently by the L3 and OPAL collaborations
at LEP2 \cite{L3_gg_ccbar,Csilling}.
The measured, positron and electron antitagged cross sections
cannot be described as a sum of direct and single-resolved
contributions, even if next-to-leading order corrections
are included \cite{Csilling}.
The measured cross section exceeds the theoretical predictions
by a large factor.
This is a new situation in comparison to the charm production
where the deficit is much smaller.

The cross section for the $e^+ e^- \rightarrow b \bar b \; X$ reaction
when both positron and electron are antitagged
can be easily estimated in the equivalent photon approximation (EPA)
as
\begin{eqnarray}
\sigma(e^+ e^- \rightarrow b \bar b X; W_{ee}) =
\int d x_A d x_B \; f_A(E_b,\theta_{max},x_A)
                    f_B(E_b,\theta_{max},x_B)
\nonumber \\
\sigma(\gamma \gamma \rightarrow b \bar b X; W) \; ,
\label{EPA}
\end{eqnarray}
where $f_A$ and $f_B$ are virtuality-integrated flux factors
of photons in the positron and electron, respectively, and
$\theta_{max}$ is the maximal angle of the
positron/electron not, to be identified by the experimental apparatus.
In the present analysis we calculate the integrated flux factors
$f_A$ and $f_B$ in a simple logarithmic approximation.
The photon-photon energy can be calculated in terms of
photon longitudinal momentum fractions $x_A$ and $x_B$ in the positron
and electron, respectively, as $W = \sqrt{x_A x_B s_{ee}}$.
It is instructive to visualize how different regions of
$W_{\gamma \gamma}$ contribute to
$\sigma(e^+ e^- \rightarrow b \bar b X; W_{ee})$.
For this purpose it is useful to transform variables
from $x_A, x_B$ to $x_F \equiv x_A - x_B$ and $W_{\gamma \gamma} = W$.
Then
\begin{eqnarray}
\sigma(e^+ e^- \rightarrow b \bar b X; W_{ee}) =
\int d x_F d W \; {\cal J} \; f_A(E_b,\theta_{max},x_A)
                              f_B(E_b,\theta_{max},x_B)
\nonumber \\
\sigma(\gamma \gamma \rightarrow b \bar b X; W) \; ,
\label{EPA_W}
\end{eqnarray}
where the Jacobian ${\cal J}$ is a simple function of kinematical
variables.



\begin{figure}

  \begin{center}
\hspace{-1.4cm}
    \includegraphics[width=5.9cm]{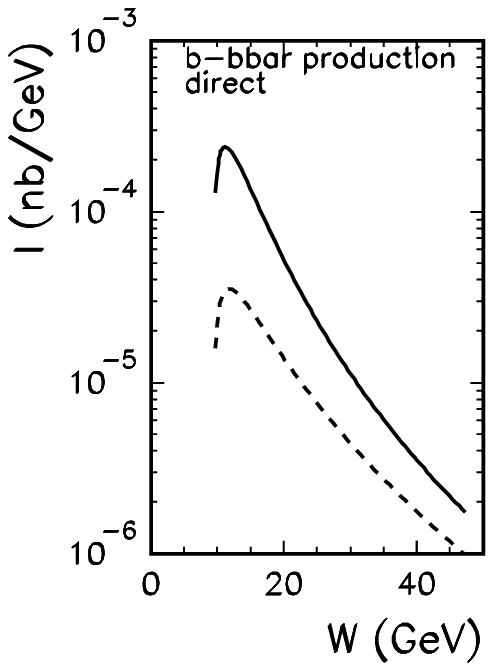}
\hspace{0.0cm}
    \includegraphics[width=5.9cm]{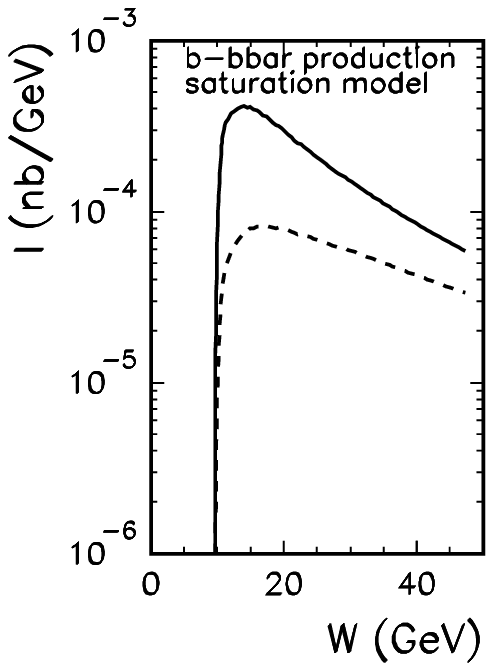}
  \end{center}
\vspace{-1.3cm}
\caption{\small
The dependence of the integrand of Eq.(\ref{EPA}) on
the photon-photon energy  $W_{\gamma \gamma}$ for $x_F$ = 0 (solid)
and $x_F$ = $\pm$ 0.5 (dashed) for the $b \bar b$ production for the direct
mechanism (left panel) and in the dipole-dipole scattering in
the Saturation Model (right panel) with the present
prescription for the energy dependence of the dipole-dipole cross
section. In this calculation $W_{ee}$ = 190 GeV.
}
\label{fig_wgg_dep}
\end{figure}


The integrand $I(x_F,W)$ of (\ref{EPA_W}) for $b \bar b$ is shown in
Fig.\ref{fig_wgg_dep} for the direct production (left panel) and for
the Saturation Model (right panel) including all contributions
considered in the present analysis. Quite a different pattern
can be observed for the two mechanisms. While for the direct production
one is sensitive mainly to low-energy photon-photon collisions,
in the Saturation Model the contributions of high-energies cannot
be neglected and one has to integrate over $W_{\gamma \gamma}$
essentially up to $W_{ee}$. This difference in $W_{\gamma \gamma}$
is due to different energy dependence of
$\sigma(\gamma \gamma \rightarrow b \bar b; W_{\gamma \gamma})$
for the different mechanisms considered, as has been discussed above.
Even in the latter case the integrated cross section is very
sensitive to the region of not too high energies $W \sim$ 20 GeV,
where the not-fully-understood threshold effects may play essential
role.


\begin{table}

\caption{\small Cross sections in pb for $e^+ e^- \rightarrow b \bar b X$
for LEP2 averaged energy $W_{ee}$ = 190 GeV.}
\begin{center}

\begin{tabular}{ |c|c|c|c|c|c|c| } 
\hline
direct & $b \bar b$ & $2 b 2 \bar b$ & SR      & sum & L3 & OPAL \\
       &  SAT-MOD   & SAT-MOD        & SAT-MOD &     &    &      \\
\hline
1.21  & 6.1-7.4 & 0.034 & 1.92  & 9.3-10.6 & 13.1 $\pm$ 2.0 $\pm$ 2.4 &
                                     14.2 $\pm$ 2.5 $\pm$ 5.0   \\
\hline

\end{tabular}

\end{center}

\end{table} 


For LEP2 averaged energy
$< W_{ee} > \approx$ 190 GeV the cross section integrated
taking into account experimental cuts is
$\sigma(e^+ e^- \rightarrow b \bar b X)$ = 6.1 pb (C = 1)
or 7.4 pb (C = 1/2) \cite{szczurek} for the dipole-dipole
SAT-MOD scattering process.
The corresponding cross section for the direct production is
$\sigma(e^+ e^- \rightarrow b \bar b X)$ = 1.2 pb.
The hadronic single-resolved contribution calculated here in
the Saturation Model is very similar
in size to that calculated in the standard collinear approach
\cite{Csilling}.
As can be seen in Table 1 the $2b 2\bar b$ contribution is
practically negligible. We have completely omitted
the double-resolved contribution which is practically negligible.
The sum of the direct, $b \bar b$ SAT-MOD,
$2 b 2 \bar b$ SAT-MOD and the hadronic single resolved SAT-MOD component
is 9.3-10.6 pb in the case when no transverse momenta cuts
on the main SAT-MOD component are included and 6.4-7.1 pb with the cuts.
These numbers should be compared to experimentally measured
$\sigma(e^+ e^- \rightarrow b \bar b X)$ = 13.1 $\pm$ 2.0 (stat)
$\pm$ 2.4 (syst) pb \cite{L3_ee_bbbar} (L3) and preliminary
$\sigma(e^+ e^- \rightarrow b \bar b X)$ = 14.2 $\pm$ 2.5 (stat)
$\pm$ 5.0 (syst) pb \cite{L3_ee_bbbar} (OPAL).
In comparison to earlier calculations in the literature,
the theoretical deficit is much smaller.
The success of the present calculation relies on the inclusion
of a few mechanisms neglected so far - in particular
the dipole-dipole contribution which, in our opinion, is not
contained in the standard collinear approach.

Up to now only the $W_{\gamma \gamma}$-integrated cross section has been
determined experimentally. This, in fact, does not allow to identify
experimentally whether the problem is in low or high $W_{\gamma \gamma}$.
In order to identify better the region where the standard collinear
approach fails it would be useful to bin the experimental cross section
in the intervals of $W_{\gamma \gamma}$ making use of a possibility
to measure $W_{vis}$ which can be related to $W_{\gamma \gamma}$
via a suitable Monte Carlo program. At present, even splitting
the cross section for $e^+ e^- \rightarrow b \bar b X$ into
$\sigma(W_{\gamma \gamma} < W_0)$ and
$\sigma(W_{\gamma \gamma} > W_0)$ for $W_0 \sim$ 20 GeV
would be useful and should shed more
light on the problem of the experimental excess of $b \bar b$ relative
to the ``standard'' QCD approach.

\subsection{Quark-antiquark correlations}

So far mainly the integrated cross section for heavy quark/antiquark
production was considered in the literature.
Only in a few cases inclusive distributions in transverse
momentum or rapidity (see e.g.\cite{KL96}) were presented.
No attempts have so far been made to analyze the final state
in more detail.
In our opinion investigating correlations between
heavy quark - heavy antiquark could be much more conclusive in
identifying the production mechanisms than the integrated cross section
or even a single variable distribution.

In principle, any correlation between two kinematical variables
of the final quark and antiquark would be of interest.
We suggest that the following final quark/antiquark momentum fractions:
\begin{eqnarray}
x_{Q} = \frac{\vec{p}_{Q}}
             { | \vec{p}_{Q} | } {\hat n}_{\gamma_1}
\;, \nonumber \\
x_{\bar Q} = \frac{\vec{p}_{\bar Q}}
             { | \vec{p}_{\bar Q} | }
 {\hat n}_{\gamma_1}
\; ,
\label{x_definition}
\end{eqnarray}
where
\begin{equation}
{\hat n_{\gamma_1}} = 
 \frac{\vec{p}_{\gamma_1}}
 { | \vec{p}_{\gamma_1} | }
\end{equation}
would be very useful to separate the different mechanisms (approaches).
In the definition above $\vec{p}_Q$ and
$\vec{p}_{\bar Q}$ are momenta of the heavy quark and antiquark,
respectively, and
$\vec{p}_{\gamma_1}$ is the momentum of the first photon,
all in the photon-photon center-of-mass frame.
By definition -1 $< x_{Q}, x_{\bar Q} <$ 1.
Similar quantities are being used at present when analyzing,
e.g., jet production at HERA to separate out resolved and direct
processes.


\begin{figure}
\vspace{-3cm}
\begin{center}
    \includegraphics[width=12cm]{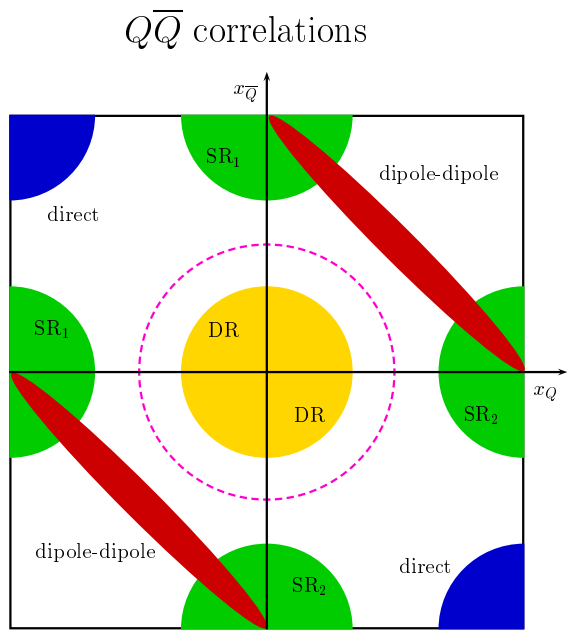}
\end{center}
\vspace{-5cm}
\caption{\small
The expected locii in $(x_{Q},x_{\bar Q})$ space
of different mechanisms considered in the present analysis.
$SR_1$ / $SR_2$ means that the first/second photon was
transformed into vector mesons and $DR$ means that each of the
photons was transformed into a vector meson.
The dashed circle is the locus corresponding to the pairs
emitted from the middle of the gluonic ladder (not discussed
in the text).
}
\label{correl}
\end{figure}


\vspace{1cm}

In Fig.\ref{correl} I present a sketch of naive expectations.
Although a precise map requires detailed calculations
for each mechanism separately, which is beyond the scope of
the present analysis, it is obvious that the separation of different
mechanisms here should be much better than for any inclusive spectra.
In the case of dipole-dipole approach (the elongated ellipses)
this would require going to the momentum representation.
The mixed representation used in the present paper is
useful only for integrated cross sections.

Experimentally, the analysis suggested would be difficult
at LEP2 because of rather limited statistics. We hope that
such an analysis will be possible at the photon-photon option at TESLA. 
At present, even localizing a few LEP2 coincidence $c \bar c$
events in the diagram $x_c$ versus $x_{\bar c}$ would be instructive.

\section{Conclusions}

There is no common consensus in the literature on detailed
understanding of the dynamics of photon-nucleon and photon-photon
collisions. In this presentation I have limited the discussion
to the production of heavy quarks simultaneously in
photon-nucleon and photon-photon collisions at high energies.
The sizeable mass of charm or bottom quarks sets
a natural low energy limit on a naive application of SAT-MOD.
Here a careful treatment of the kinematical threshold is required.

We have started the analysis from (real) photon-nucleon scattering,
which is very close to the domain of 
SAT-MOD as formulated in \cite{GW98}.
If the kinematical threshold corrections are
included, SAT-MOD gives similar results as the standard
collinear approach for both charm and bottom production.
We have estimated the VDM contribution
to the heavy quark production.

The second part of the present analysis has been devoted to
real photon - real photon collisions.
For the first time in the literature we have estimated
the cross section for the production of $2 c 2 \bar c$ final state.
 We have found that this component
constitutes up to 10-15 \% of the inclusive charm production
at high energies and is negligible for the bottom production.
We have shown how to generalize SM
to the case when one of the photons fluctuates into
light vector mesons. It was found that this component yields
a significant correction of about 30-40 \% for inclusive
charm production and 15-20 \% for bottom production.
We have shown that the double resolved component, when both
photons fluctuate into light vector mesons, is nonnegligible only
at very high energies, both for the charm and bottom production.

I have shown that the production of $c \bar c$ pairs
(the same is true for $b \bar b$) is not completely
of perturbative character and involves both short- and large-size
contributions. The latter as nonperturbative are unavoidably
subjected to some modeling. Present experimental statistics do not
allow extraction of cross sections for the
$\gamma \gamma \rightarrow b \bar b$ reaction and therefore
it is not clear where the observed deficit resides.
It is not excluded that the apparent deficit of bottom quarks
may reside at photon-photon energies close to threshold.
This is a region where the underlying physics has never carefully
been studied.

Finally I have discussed a possibility to distinguish experimentally
the different mechanisms discussed in the present paper by measuring
heavy quark - antiquark correlations. This suggestion requires,
however, further detailed studies of the Monte Carlo type, including
experimental possibilities and limitations.

The present calculation is not fully consistent as far as parameters
are considered. I have taken SAT-MOD parameters from Ref.\cite{GW98},
where only dipole (quark-antiquark) component was fitted to
the total cross section for $\gamma^* p$ scattering and added
the resolved photon in addition.
In a consistent treatment one should include in addition
the resolved photon component in the fit to the HERA
structure function data.
Such an analysis has been completed only recently \cite{PS03}.
%


\end{document}